\documentclass[english]{revtex4}
\usepackage{times}
\usepackage[T1]{fontenc}
\usepackage[latin1]{inputenc}
\usepackage{amsmath}
\usepackage{graphicx}
\usepackage{amssymb}
\makeatletter
\usepackage{babel}
\makeatother
\begin{document}

\title{Inflationary Cosmology Connecting Dark Energy and Dark Matter}

\author{Daniel J.~H.~Chung}\email{danielchung@wisc.edu}
\author{Lisa L. Everett}\email{leverett@wisc.edu}
\affiliation{Department of Physics, University of Wisconsin,
  Madison, WI 53706, USA}

\author{Konstantin T.~Matchev}\email{matchev@phys.ufl.edu}
\affiliation{Institute for Fundamental Theory, Physics Department,
  University of Florida, Gainesville, FL 32611, USA}

\date{April 23, 2007}
\rightline{UFIFT-HEP-07-5}
\rightline{MADPH-07-1487}

\begin{abstract}
Kination dominated quintessence models of dark
energy have the intriguing feature that the relic abundance of
thermal cold dark matter can be significantly enhanced compared to the
predictions from standard cosmology.  Previous treatments of such models do not include a realistic embedding of inflationary initial conditions.  We
remedy this situation by constructing a viable inflationary model in
which the inflaton and quintessence field are the same scalar degree
of freedom.  Kination domination is achieved after inflation through a
strong push or ``kick" of the inflaton, and sufficient reheating can
be achieved depending on model parameters.  This allows us to 
explore both model-dependent and model-independent cosmological
predictions of this scenario.  We find that measurements of the B-mode
CMB polarization can rule out this class of scenarios almost model
independently.  We also discuss other experimentally accessible
signatures for this class of models.
\end{abstract}
\maketitle

\section{Introduction}

The discovery that the universe is dominated by dark energy strongly
suggests that the predictions of standard cosmology should be
reevaluated.  An intriguing possible explanation of the nature of dark
energy arises within the quintessence paradigm, in which the dark
energy takes the form of a slowly evolving scalar field (see e.g. \cite{Caldwell:1997ii,Wetterich:1987fk,Peebles:1987ek,Freese:1986dd,bertolami}).  This
scenario is extremely difficult (if not impossible) to test directly
in collider experiments, since quintessence models generically require
the quintessence field to have gravitationally suppressed interactions
with the fields of the Standard Model (SM).

However, if the dark energy is in the form of quintessence, the
presence of the quintessence field can modify the cosmological
evolution and lead to significant departures from standard cosmology.
A striking example is the possible interconnection of dark matter and
dark energy within quintessence scenarios
\cite{Salati:2002md,Rosati:2003yw,Profumo:2003hq,Pallis:2005hm,Barenboim:2006jh}.
As first pointed out by Salati \cite{Salati:2002md}, the freeze-out of
thermal relics can be strongly enhanced in scenarios in which the
energy density is dominated by the kinetic energy of the quintessence
field (kination domination) during the time of freeze-out, but dilutes
away by the time of big bang nucleosynthesis (BBN).  (Related
scenarios were also suggested before by
\cite{Kamionkowski:1990ni,Barrow:1982ei}.)  Such kination dominated
freeze-out scenarios are then consistent with standard cosmology and
predict that the standard relic abundance computed from the parameters
extracted from collider measurements will be mismatched from the relic
abundance deduced by observational cosmology.  This has implications
for TeV physics models with thermal dark matter candidates
(e.g. models with low energy supersymmetry such as the minimal
supersymmetric extension of the SM (MSSM), technicolor models, models
with large/warped extra dimensions, or certain classes of little Higgs
models), which will be probed at the LHC and other experiments in the
foreseeable future.

In most of the previous discussions of this class of scenarios
\cite{Salati:2002md,Rosati:2003yw,Profumo:2003hq,Pallis:2005hm}, the initial condition that the
quintessence field kinetic energy density is the dominant component
was put in as an ansatz, without a complete picture of the inflationary dynamics. In this paper, we
address this issue by constructing an inflationary scenario that dynamically leads to a
kination dominated quintessence period.  This yields robust
predictions which can be used to experimentally support or rule out
this class of scenarios.

To construct viable inflationary models which lead to kination
domination, the following constraints must be satisfied:
\begin{enumerate}
\item The energy density in a coherent quintessence field must dominate over
  radiation after the end of inflation.
\item The quintessence field must be kinetic energy dominated.
\item The inflaton potential must satisfy the usual requirements of
  a sufficient number of efoldings, the right amplitude of density
  perturbations, and a nearly scale invariant spectral
  index (with a slight preference for a red spectrum \cite{Spergel:2006hy}).
\item There must not be too much reheating in the phase transition at the end of
  inflation (when the quasi-de Sitter phase ends), such that the ratio of
  the kination energy density to the radiation energy density can be
  large.
\end{enumerate}
To satisfy these constraints, we build a model in which the
quintessence field is the inflaton field $\Phi$, which has 
sufficiently weak couplings to the SM fields at the end of inflation
such that most of the energy density responsible for inflation gets
converted to $\Phi$ coherent kinetic energy in a runaway
potential. Radiation domination is achieved because the coherent
kinetic energy dilutes as $1/a^{6}$, while the suppressed radiation
produced at the end of inflation gets relatively amplified since it
dilutes as $1/a^{4}$.  The choice of the inflaton as the quintessence
field in kination domination is natural in this framework, since the inflaton possesses
the required qualities for the kination domination construction:
energy dominance and coherence.  Other quintessence models have been
constructed in which the inflaton is the same field as the
quintessence field (see e.g. \cite{Dimopoulos:2001ix,Barenboim:2005np,Barenboim:2006rx,Bertolami:2006zg,Sanchez:2006ah}), but we treat the reheating more precisely and derive new predictions in the context of kination-dominated quintessence models relevant for dark matter.  The work \cite{Barenboim:2006jh} considers the connection between dark energy and dark matter, but it differs from the present treatment in that their scenario is model specific and has multiple inflationary phases. 

In addition to the prediction that the relic abundance
inferred from collider measurements can measurably disagree with the cosmologically
inferred dark matter relic abundance (in the context of a thermal
freeze-out scenario), many experimentally accessible cosmological
predictions can be made by embedding the kination domination scenario
in inflationary models. The most {\em model independent} nontrivial
prediction is the absence of a measurable tensor perturbation induced
B-mode CMB polarization.  Hence, the next generation of CMB
experiments can falsify this class of models.

Other predictions include a shift in the peak of the gravity wave
signal from the electroweak phase transition, shifts in the verifiable
leptogenesis/baryogenesis scenarios whose out of equilibrium
ingredient is furnished by $H$, and shifts in any residual annihilation
effects.  In general, the indirect dark matter detection signals are enhanced in this scenario \cite{Profumo:2004ty}.
In particular, it would be interesting to see whether one can explain the positron excess as observed by HEAT and other experiments near 7 GeV within this scenario \cite{gordyinprogress}, 
 since most attempts to explain 
this excess in terms of dark matter annihilations require sufficiently large cross sections that the relic abundance would be negligible with standard cosmological assumptions.

The order of presentation will be as follows.  We begin by
analyzing both analytically and numerically the constraints on
quintessence potentials to achieve a period of kination domination,  
and look for equation of state signatures.  In Section \ref{sec:inflation}, we construct a
class of inflationary models embedding the kination domination
scenario and discuss a robust prediction that can observationally
falsify this class of models.  In Section \ref{sec:otherpredictions},
we discuss other observationally accessible cosmological predictions
which may corroborate this class of scenarios.  We then summarize and
conclude.  In the Appendix, we provide the details of the particle
production computation for the unusual reheating scenario associated
with this class of inflationary scenarios.

Throughout this paper, we use the convention of $M_p$ to denote the
reduced Planck mass of approximately $2.4 \times 10^{18}$ GeV.
 
\section{Mapping Kinetic Behavior to Quintessence Potentials}
The standard procedure in constructing scenarios of quintessence
dynamics is to focus on potentials, which is a sensible approach since
a negative equation of state requires potential energy domination.
However, when considering quintessence dynamics with a period of
kination domination, it is more natural to focus on the behavior of
the field velocity, as this is the quantity which characterizes the
energy density of the quintessence field.  Here we develop a formalism
to map the desired behavior of the kinetic energy to a class of scalar
potentials.  Our results show that the allowed quintessence potentials
are not severely restricted by the requirement of a period of kination
domination, since the time at which the equation of state is close to
$-1$ is necessarily much later than the time of dark matter freeze-out
due to the strong constraints from BBN.  In what follows, we will not
restrict ourselves to ``tracker" models, in order to separate the
difficulties of constructing good trackers from the constraints
imposed by kination domination. It is worth noting that kination
dominated initial conditions can be obtained in tracking potentials,
as discussed in \cite{Profumo:2003hq,Catena:2004ba}, although clearly
there will be constraints depending on the exact form of the
potential.

To carry out the engineering of quintessence potentials which match the
desired kinetic term histories, we begin with the familiar equations of motion
for the quintessence field $\Phi$ and the Friedmann-Robertson-Walker (FRW) scale factor $H\equiv
\dot{a}/a$:
\begin{equation}
\ddot{\Phi}+3H\dot{\Phi}+V'(\Phi)=0
\label{quinteom}
\end{equation}
\begin{equation}
H=\sqrt{\frac{1}{3M_{p}^2}\left (\frac{1}{2}\dot{\Phi}^2+V(\Phi) +\rho_{RM}\right )},
\label{frw}
\end{equation}
in which  $\rho_{RM}\equiv\rho_{R}+\rho_{M}$ corresponds to the energy
densities of radiation and matter.  Using the following definitions, 
\begin{equation}
q(\Phi)\equiv\ln a^{3}(\Phi)\label{eq:qasafuncofa}\end{equation}
\begin{equation}
f(\Phi)\equiv\dot{\Phi}(t)\end{equation}
\begin{equation}
\gamma\equiv\frac{1}{3M_{p}^{2}}, \end{equation}
we rewrite Eqs.~(\ref{quinteom}) and (\ref{frw}) as follows:
 \begin{equation}
q'(\Phi)=-\left [\frac{V'(\Phi)}{f^{2}(\Phi)}+\frac{f'(\Phi)}{f(\Phi)} \right ]\label{eq:eom}\end{equation}
\begin{equation}
q'(\Phi)=\frac{3\sqrt{\gamma}}{f(\Phi)}\sqrt{\frac{1}{2}f^{2}(\Phi)+V(\Phi)+\rho_{RM}(q(\Phi))}.\label{eq:hubble}\end{equation}
In the above, we have assumed that $\dot{\Phi}$
is a single valued function of $\Phi$ (which would exclude e.g.  the situation of oscillations). For the scenarios considered here this is
typically not a good assumption throughout the entire evolution, since the kinetic energy is usually large enough that $\Phi$ can overshoot the minimum and eventually hit the {}``other side'' of the potential before the quintessence equation of state reaches $-1$.  However, the condition on the quintessence potential during kination domination does not change even if this behavior is properly accounted for since the {}``bounce'' occurs long after the kination period is over. 

We can use Eqs.~(\ref{eq:eom})--(\ref{eq:hubble}) to solve for $q(\Phi)$ and $f(\Phi)$  if $V(\Phi)$ is given; alternatively, we can solve for $V(\Phi)$ and $q(\Phi)$  if the field velocity function $f(\Phi)$ ({\it i.e.} the kinetic energy) is specified. For generic potentials, there is no obvious obstruction to achieving a period of kination domination which leads to a period of potential energy domination, as $f(\Phi)$ can be chosen to vanish as $\Phi$ approaches a particular asymptotic
value, and the Hubble friction naturally allows for $f(\Phi)$ to vanish without $V$ also vanishing. 

In this paper, we restrict our attention to the specific
case of the quintessence field velocity function with the initial behavior (during the time
relevant for dark matter freeze-out) of \begin{equation}
f\approx f_{i}e^{q_{i}-q}\propto\frac{1}{a^{3}},\label{eq:kinationdef}\end{equation}
in which $f_i$ and $q_i$ denote initial values of the functions $f(\Phi)$ and $q(\Phi)$.  This corresponds to the kination regime in which the universe is driven by quintessence
kinetic energy,  with $V'/f^{2}$ playing a subdominant role compared to $f'/f$ in the equation of motion for $\Phi$ ({\it i.e.}, the force term is subdominant to the Hubble friction term). The reason of course for this restriction is that kination domination gives
rise to an energy density which dilutes as $1/a^{6}$, which allows the quintessence energy to be important during DM freeze-out but disappear by the time of BBN as required by phenomenology. 

Let us now study what a potential that leads to the kination behavior
of Eq.~(\ref{eq:kinationdef}) looks like by constructing it using
Eqs.~(\ref{eq:eom}) and (\ref{eq:hubble}). To obtain a closed form
solution, we take the ansatz of neglecting $V'/f^{2}$, in which case we 
obtain\begin{equation}
q=q_{i}-\ln\frac{f(\Phi)}{f_{i}}\label{eq:case1}\end{equation}
\begin{equation}
V(\Phi)=\frac{-1}{2}f^{2}(\Phi)-\rho_{RM}(q(\Phi))+\frac{1}{9\gamma}[f'(\Phi)]^{2}.\label{eq:potential}\end{equation}
One possible procedure is then as follows: choose any desired behavior of $f$ and obtain using
Eqs. (\ref{eq:case1}) and (\ref{eq:potential}) a corresponding expansion
history (specified by $q$) and the potential $V$, then check that
$V'/f^{2}$ can be neglected in the equation of motion. The validity
of neglecting $V'/f^{2}$ can be rewritten as \begin{equation}
\left |-1+\frac{1}{f^{2}}\frac{d\rho_{RM}}{dq}+\frac{2}{9\gamma}\frac{f''}{f} \right |\ll1.\label{eq:neglectcase1}\end{equation}
One trivial way to make $V'/f^{2}$ negligible is to have $V$ much
smaller than any other energy component and to have $V$ be a smooth
function.  Another way to satisfy Eq.~(\ref{eq:neglectcase1}) is to set
its left hand side 
equal to an arbitrary
small function $h(\Phi)\ll 1$ and solve for $f(\Phi)$.  For example,
if the universe is radiation dominated, we can write 
\begin{equation}\rho_{RM}(q) =
\rho_{i}\left (\frac{a_{i}}{a} \right)^4 
=\rho_{i}e^{\frac{4}{3}(q_{i}-q)}
\end{equation}
 to obtain the following equation:
\begin{equation}
f''(\Phi) - 6 \gamma \left(\frac{f(\Phi)}{f_i} \right)^{1/3} \kappa f_i - 
\frac{9 \gamma}{2} (1+ h(\Phi)) f(\Phi)=0,
\label{eq:secondorderform}
\end{equation}
in which 
\begin{equation}
\kappa= \rho_i/fi^2 \approx \frac{5\times 10^{-7}}{\eta_\Phi},
\end{equation}
with $\eta_\Phi \equiv \rho_{\Phi}/\rho_{\gamma} |_{T=1\,{\rm MeV}}$, and $f_i \equiv f(\Phi_i) \approx 6\times 10^3 \sqrt{\eta_\Phi} \mbox{GeV}^2$ (the initial field velocity) as defined at a temperature of approximately 1 GeV. Note that 
since $h(\Phi) \ll 1$ is arbitrary, nearly any smoothly
varying $f(\Phi)$ can be obtained, which in turn means any smoothly varying shape for
the potential can be obtained even with the constraint of Eq.~(\ref{eq:kinationdef}).
(This corresponds of course to the case in which the $V$ energy density is
negligible during kination domination.) To obtain intuition, if we
set $h(\Phi)=h_0=\mbox{constant}\ll 1$ and define $y\equiv f(\Phi)/f_i$ and $x=
\Phi \sqrt{\gamma}$,  we can rewrite
Eq.~(\ref{eq:secondorderform}) in terms of a first order equation:
\begin{equation}
\frac{dy}{dx} = - \sqrt{ 9\kappa y^{4/3} + \frac{9}{2} (1+h_0) y^2 +C},
\label{eq:frewritten}
\end{equation}
in which $C$ is an integration constant.   Note that we have chosen the negative squareroot
to coincide with the convention in this paper that $\Phi$ is moving in
the positive direction during kination domination (recall $y$ is decreasing by construction).

Although Eq.~(\ref{eq:frewritten}) can be used to write the solution
as a single integral, the phase space of solutions can also be visualized
by examining this equation.
As long as $C \geq 0$, $y$
moves towards $y=0$ ($y \sim 2 \times 10^{-10}$ at the time of  
BBN). The existence of a
large class of solutions given by $C \geq 0$ is not surprising given that
the Hubble expansion generically provides friction for the field
velocity $f$.  The time it takes for the field to achieve a final
velocity $f_f$ can be written as an integral:
\begin{equation}
\Delta t = \frac{1}{f_i \sqrt{\gamma}} \int_1^{f_f/f_i} \frac{dy}{y
  \frac{dy}{dx}}.
\label{eq:timeinterval}
\end{equation}
As a consistency check,  it can be seen from Eq.~(\ref{eq:frewritten})
that generically it takes an infinitely long time to achieve $f_f=0$ (an infinite time is necessary since we are engineering $f\propto
a^{-3}$).  Although a larger $C$ seems to correspond to a shorter time
to achieve the desired $f_f$, it also implies 
that the potential is larger, which places severe
constraints on $C$.  More explicitly, it can be shown that the potential of
Eq.~(\ref{eq:potential}) implied by Eq.~(\ref{eq:frewritten}) is of the form
\begin{equation}
V(\Phi)= \left(\frac{2 C}{9} +h_0 y^2(\sqrt{\gamma}\Phi)\right)
\frac{f_i^2}{2},
\label{eq:explicitpotential}
\end{equation}
where $C$ is then explicitly seen to control the cosmological
constant.  (If $C<0$, the cosmological constant will be negative, which will cause the universe to eventually contract.  This is consistent with the $C\geq 0$ condition described above.)   Note that the potential is independent of $\kappa$ except
through $y(x)$.  It can be seen from this expression
that if $h(\Phi)$ were not a constant $h_0$, a richer shape of
potentials can easily be attained ({\it i.e.}, it is easy to show that the
$y$ dependent term generalizes to $f_i^2 \int dy y h$ when $h$ is not
a constant).  Furthermore, we see the condition
that $V(\Phi)$ be a subdominant energy component to $f^2/2$ implies
that $C \ll 1$.  Specifically, if  $C f_i^2/9$ is identified as
the cosmological constant that persists to today, one would impose the
constraint $C \lesssim 9 \Omega_\Lambda \rho_C/f_i^2 \approx 6\times
10^{-54}/\eta_\Phi$.   If we insist on this form of the
potential until BBN, we have a bound of
\begin{equation}
C  \ll \frac{10^{-18}}{\eta_\Phi}
\end{equation}
even if this piece of the potential were
piecewise glued to other functional forms for the potential once the evolution has persisted past
the kination dominated period.
Since $C$ is generally required to be small during kination
domination, we can now solve Eq.~(\ref{eq:frewritten}) exactly neglecting
$C$, which leads to the expression
\begin{equation}
y(x)= \frac{1}{64 (1+h_0)^{3/2}} \exp(- 3 \sqrt{1+ h_0}
(x-C_2)/\sqrt{2}) \left(1- 8\kappa \exp[\sqrt{2 (1+h_0)}(x-C_2)] \right)^3,
\label{eq:explicitsolforpotential}
\end{equation}
in which $C_2$ is another integration constant that is specified by
$y(x_i)=1$:
\begin{equation}
C_2 = x_i+ \frac{\sqrt{2}}{\sqrt{1+h_0}} \ln\left[2(\sqrt{1+h_0}+
    \sqrt{1+h_0+2 \kappa}) \right]
\end{equation}
As expected, initially the radiation energy density encoded in
$\kappa$ is unimportant, but once the radiation catches up with the
quintessence kinetic energy, the potential 
must compensate to (artificially) maintain the 
$1/a^6$ behavior for the quintessence energy density while keeping the
left hand side of Eq.~(\ref{eq:neglectcase1}), $h(\Phi)$, a constant.
Once again, it is important to note that the shape of the potential
dictated by Eq.~(\ref{eq:explicitsolforpotential}) is not fundamental
because this solution is only valid for a constant $h(\Phi)$.

The main lesson from the discussion thus far is that potentials can be
chosen to not interfere with $1/a^6$ behavior of the quintessence
energy density even past the point in which the radiation starts to
dominate.  (As discussed earlier, this follows from the fact that the
solution to the $\Phi$ equation of motion with $V'(\Phi)=0$ is
$\dot{\Phi}\propto a^{-3}$.)   To obtain a potential that is manifestly independent of
$\kappa$, we can choose implicitly a non-constant $h(\Phi)$.  For
example, inspired by Eq.~(\ref{eq:explicitsolforpotential}), we can
choose the potential to be 
\begin{equation}
V= \Omega_\Lambda \rho_c [1+b \cosh^2(\lambda \Phi)], 
\label{eq:potential1}
\end{equation}
in which 
\begin{equation}
\lambda\equiv \sqrt{\frac{9\gamma}{2}(1+b)},
\end{equation}
with $b\ll 1$.  This form of the potential results in $V'(\Phi)$
being unimportant during the $\Phi$ evolution until the temperature reaches
\begin{equation}
T\sim10^{-10}b^{2/9}(1+b)^{1/9}\exp\left(-10\sqrt{1+b}+\frac{4n}{9}\right)\eta_{\Phi}^{-\frac{1}{9}-\frac{2}{3}\sqrt{1+b}}\mbox{ GeV},
\end{equation}
where the initial value of $\Phi$ was parametrized as $-n/\lambda$.
The values $n=30$, $b=10^{-6}$, and $\eta_\Phi=1$ result in a temperature during matter domination but close to the matter-radiation equality ({\it i.e.}, long after the end of BBN).  Therefore, this potential provides the desired kination behavior.  We will be using this form of the potential for the rest of the paper.

To check the stability of the background scalar field solutions, we
set $\Phi=\phi_{B}(t)+a^{-3/2}\delta\tilde{\phi}(t,\vec{x})$, in which
$\phi_{B}(t)$ is the background solution, and note that upon
neglecting the metric fluctuations to a leading order approximation,
the equations of motion for the field fluctuations take the
form:\begin{equation}
\delta\tilde{\phi}-\frac{1}{a^{2}}(\partial_{i}\delta\tilde{\phi})^{2}+[V''(\phi_{B})+\frac{9}{4}\gamma
P]\delta\tilde{\phi}=0,\end{equation} in which $P$ is the pressure of
an ideal fluid (our stress tensor approximation).  The condition for
stability on all length scales is
\begin{equation}V''(\phi_{B})+\frac{9}{4}\gamma P>0.\end{equation}
Therefore, if $V''(\phi_{B})>0$ (which is true for example in the case
of constant $h(\Phi)$ in Eq.~(\ref{eq:explicitpotential}) if $h_0>0$
or in the case of Eq.~(\ref{eq:potential1}) if $b>0$),
the solution will always be stable for the case of $P>0$ relevant for
kination domination scenarios.

The viability of the quintessence picture also requires that the background solution
tends toward a potential energy dominated regime with
%
$V(\Phi)\sim\Omega_{\Lambda}\rho_{c}$,
in which $\rho_{c}$ is the critical energy density today. Since the equation of state can be written as \begin{eqnarray} w & = & \frac{1-2V/f^{2}+\frac{2}{3}\rho_{R}/f^{2}}{1+2V/f^{2}+2(\rho_{M}+\rho_{R})/f^{2}},\end{eqnarray}
it is straightforward to see that as long as $V$ asymptotically dominates the energy density as the
universe expands, the kination scenario naturally leads to the desired
late time quintessence behavior of $w\rightarrow-1$. 
The phenomenological requirement is that $w\approx-1$ by redshift of about $z\approx1$.

A final issue to address is the underlying dynamics which leads to the
required initial conditions to achieve the desired $f_{i}$.  We will
return to this question later and will discuss interesting
observational consequences, but first analyze an explicit example
numerically to illustrate the possible resulting equation of state.

In solving the quintessential cosmology specified by
Eq. (\ref{eq:potential1}) numerically, we choose the following representative parameters:
 $\Omega_{\Lambda}=0.72$, $\Omega_{M}=0.28$
(including baryons), $\Omega_{R}=4.6\times10^{-5}$, and
$H_{0}=73\textrm{ km/s/Mpc}$.  We further set $b=10^{-6}$,
$\lambda\Phi(t_{\textrm{tinitial}})=\{-30,-20,-10\}$,
$\eta_{\Phi}\equiv \rho_\Phi/\rho_\gamma|_{T=1MeV}=0.5$, and
$\lambda\dot{\Phi}(t_{\textrm{initial}})>0$ (the magnitude is fixed by
$\eta_{\Phi}$). We also take the effective degrees of freedom to
evolve approximately as\begin{equation} g_{*}(T)\approx
g_{*S}(T)\approx\left\{ \begin{array}{cc} 90 & T>1\textrm{ GeV}\\ 60 &
1\textrm{ GeV}>T>0.1\textrm{ GeV}\\ 10.75 & 0.1\textrm{
GeV}>T>10^{-4}\textrm{ GeV}\\ 3.36 & 10^{-4}\textrm{
GeV}>T.\end{array}\right.\label{eq:stepfunctions}\end{equation}
Although the relation $g_{*}(T)\approx g_{*S}(T)$ breaks down at late
times, these subtleties do not affect the main results for kination
domination. In evolving the coupled differential equations for $\{
a(t),\Phi(t)\}$ as given in Eq.~(\ref{quinteom}) and Eq.~(\ref{frw}),
we take \begin{equation}
\rho_{R}(t)=\rho_{R}(\textrm{today})\left(\frac{g_{*}(T_{\textrm{today}})}{g_{*}(T_{\textrm{approx}}(t))}\right)^{1/3}\left(\frac{a(\textrm{today})}{a(t)}\right)^{4}\label{eq:radiation}\end{equation}
and\begin{equation}
\rho_{M}(t)=\rho_{M}(\textrm{today})\left(\frac{a(\textrm{today})}{a(t)}\right)^{3}.\end{equation}
This neglects the fact that $\rho_{M}$ has a nontrivial time behavior
due to annihilations; however, the annihilation corrections are
negligible by the time $\rho_{M}$ becomes a significant component of
the energy density.  We are also using an approximate temperature, \begin{equation}
T_{\textrm{approx}}(t)\equiv
T_{\textrm{today}}\frac{a_{\textrm{today}}}{a(t)} \end{equation} in $g_{*}$ in
Eq.~(\ref{eq:radiation}).  This should be sufficiently accurate at the
current level of approximation, since $g_{*}$ is a very flat function
except in critical regions of Eq. (\ref{eq:stepfunctions}) where the
transition temperatures are only accurate to an order of magnitude.
\begin{figure}
\begin{center}\includegraphics[scale=0.5]{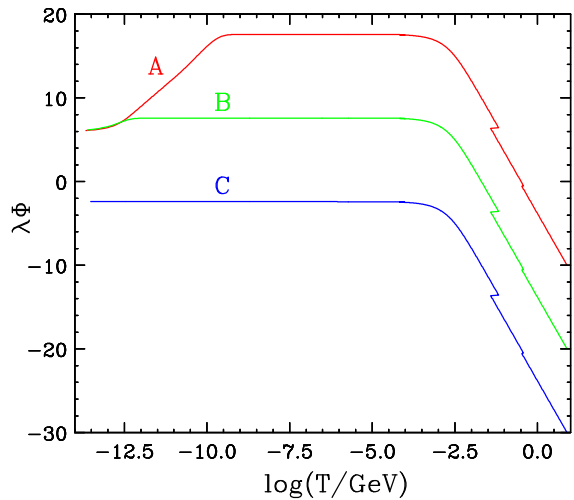}\end{center}
\caption{\label{fig:phiasfuncofT}$\Phi$ as a function of $T$. The evolution for $\lambda\Phi(t_{\textrm{tinitial}})=\{-10,-20,-30\}$ is depicted by the curves denoted by Models A, B, and C, respectively.  The little jump in temperature at $T\approx10^{-1}\textrm{ GeV}$ corresponds to the change in $g_{*}$ at that temperature; a much smaller jump can be seen at $T\approx1$ GeV. }
\end{figure}
\begin{figure}
\begin{center}
\includegraphics[scale=0.5]{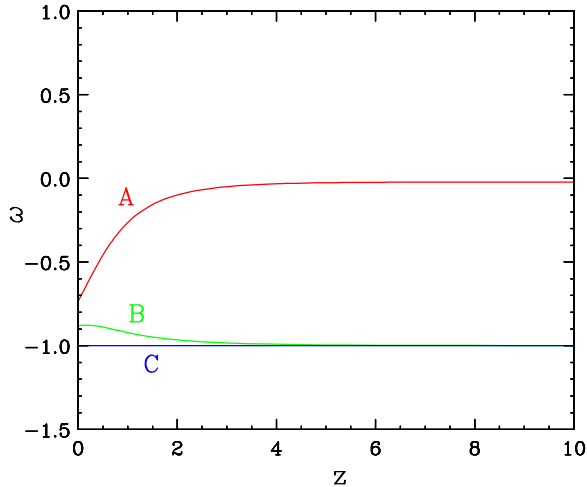}
\end{center}
\caption{\label{fig:deeos1}The dark energy equation of state $w$ as a function
of redshift $z$ for the same cases shown in Fig. \ref{fig:phiasfuncofT}.}
\end{figure}
\begin{figure}
\begin{center}\includegraphics[scale=0.5]{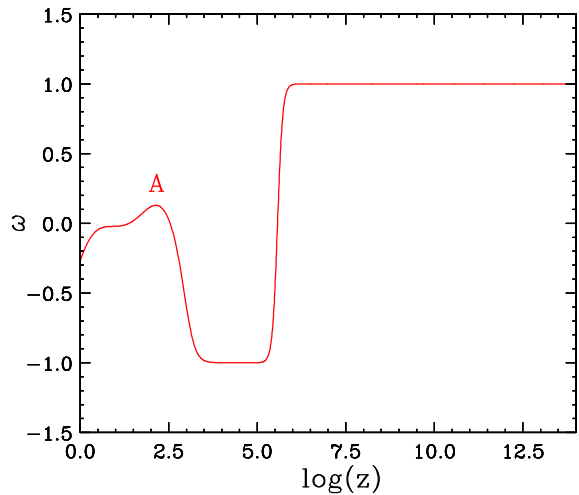}\end{center}
\caption{\label{fig:phi10case} The equation of state $w$ of the $\Phi$ component as a
function of redshift $z$ for large redshifts for Model A ($\lambda\Phi(t_{\textrm{tinitial}})=-10$). }
\end{figure}

The resulting evolution of $\Phi$ is shown in
Fig.~\ref{fig:phiasfuncofT} as a function of temperature. For
$\lambda\Phi(t_{\textrm{tinitial}})=-30$, which we denote as Model C,  the field does not reach the
origin (the minimum of the potential) by today. However, when the
initial value of $\lambda\Phi$ is less negative, the field overshoots
the minimum and climbs up the potential for $\Phi>0$, and
eventually falls back towards the origin. An inspection of
Fig.~\ref{fig:deeos1}, which shows the dark energy equation of state
as a function of redshift, demonstrates that acceptable phenomenology
can be obtained for the {}``overshooting'' case of
$\lambda\Phi(t_{\textrm{tinitial}})=-20$ (Model B), but not for
$\lambda\Phi(t_{\textrm{tinitial}})=-10$ (Model A), where there is still too
much kinetic energy by redshifts of less than one.  In fact, 
the equation of state
$w$ for Model A actually reaches $-1$ during the {}``bounce'' and then a time
period in which $w=0$ is sustained until the kinetic energy is finally
dissipated.  This nontrivial evolution of the equation of state for
larger redshifts than those shown in Fig.~\ref{fig:deeos1} can be seen
in Fig.~\ref{fig:phi10case}.  Furthermore, it can also be shown that
the scalar field energy density always dominates in Model A,
which makes such
scenarios incompatible with the successes of CDM large scale structure
phenomenology. Clearly, even with exactly same potential function
$V(\Phi)$ and initial value of quintessence kinetic energy, the dark
energy phenomenology crucially depends on the initial value of $\Phi$. 

In all of these cases, the initial condition of $|\lambda\Phi(t_{\textrm{initial}})|>10$ corresponds to
situations in which the initial displacement is trans-Planckian.  This is not attractive from an effective field theory point of view since the theory then is sensitive to all powers of the field
operator; however, such initial field values are typical in
quintessence scenarios \cite{Binetruy:2000mh}.
Indeed, although quintessence is
generically not attractive as an effective field theory candidate, its
classical dynamics may still be useful to parameterize vacuum
dynamics,  since effective field theory has grossly failed as far as understanding the 
vacuum structure is concerned, {\it i.e.}, in giving a plausible explanation
for the cosmological constant problem.  In fact,  we have implicitly been assuming throughout this paper that the cosmological constant problem has been solved by some unknown mechanism which leaves the quintessence dynamics responsible only for the nontrivial
vacuum energy dynamics.

The main lesson to be learned from this exercise is that in viable kination
scenarios, the dark energy equation of state can exhibit a wide range
of behavior both because of the shape of the potential and because of
the unknown initial conditions.  However, a typical behavior more
likely in quintessence scenarios with a period of kination domination
than in generic quintessence models is the ``bounce'' or turnaround
behavior, in which $\Phi$ bounces off the ``other'' end of the
potential barrier and slowly rolls toward the minimum. With one
bounce, the behavior of the Model A case in Fig.~\ref{fig:deeos1} in
which $dw/dz\lesssim0$ today is typical in viable models. The
complement, $dw/dz>0$, is less generic for large initial kinetic
energies, but may be possible in models in which the kinetic energy
during the kination period is sufficiently small and tuned
appropriately. Multiple bounce scenarios may also be possible in
kination scenarios, but they require a special conspiracy between the
shape of the potential and initial conditions, as exemplified by the
unsuccessful case shown in Fig.  \ref{fig:deeos1}.

In the present model, we can compute the minimum initial
displacement $|\lambda\Phi(t_{\textrm{tinitial}})|$ to avoid the
turnaround behavior. The total displacement of $\Phi$ for the case
in which $V'(\Phi)$ plays a negligible role in the equation of motion
is 
\begin{equation}
\Delta\Phi\approx\sqrt{6}M_{p}\ln\left
[\frac{a(t_{\textrm{BBN}})}{a(t_{i})}\right
]+3\sqrt{\eta_{\Phi}}M_{p},\label{displ}
\end{equation} in which $a(t_{i})$ is the scale factor at the time of
the initial position of $\Phi$, and the second term on the right hand
side of Eq.~(\ref{displ}) arises from the movement of $\Phi$ after
BBN. For typical situations in which
$a(t_{\textrm{BBN}})/a(t_{i})=a(t_{\textrm{BBN}})/a(t_{F})\gtrsim10^{3}$,
where $a(t_{F})$ is the scale factor at the time of freeze-out, we
find\begin{equation}
\Delta\Phi\gtrsim[17+3\sqrt{\eta_{\Phi}}]M_{p}.\end{equation}
This result implies that most of the displacement of $\Phi$ occurs
during the kination period and not after BBN, and that the numerical
result of Fig.~\ref{fig:phiasfuncofT} is reasonable. In addition, the
field displacement is generically transPlanckian (as discussed
previously).

Having discussed the consequences of having large quintessence kinetic
energy for late time cosmology, we now consider how such initial
conditions can be established within the context of inflationary cosmology.

\section{\label{sec:inflation} Inflationary Cosmology}

In this section, we address the issue of how kination domination might
be achieved within theories of cosmological initial conditions such as
inflation.  To demonstrate an existence proof of a viable scenario, we
construct an inflationary model in which a single scalar field plays
the role of both the inflaton and the quintessence field.  We will not
only show that kination domination can be achieved after inflation,
but that sufficient reheating can be achieved as well.  The main
benefits of embedding quintessence within an inflationary scenario are
the resulting cosmological predictions which can be experimentally
verified or falsified.  We will see that a robust prediction of the kination
scenario relevant for dark matter abundance is an absence of a
measurable B-mode CMB polarization signal. Other predictions will
include those connected to the fact that the Hubble expansion rate
during kination domination is different from that within standard cosmological scenarios even
when the radiation is in equilibrium.

The basic idea we implement to construct this inflationary model is
the assumption that the inflaton receives a kick (or a strong push) at
the end of inflation to achieve kination domination.  Indeed, since
slow roll inflationary models already require a coherent homogeneous
scalar field to dominate the energy density in the universe, the
inflaton is an ideal candidate field to be converted into a
quintessence driving kination dominated universe. By comparing a
multi-scalar field system in which the inflaton energy density
converts efficiently into a coherent scalar field kinetic energy of
another field direction, it is straightforward to see that the only
significant simplification being made by considering one scalar field
is in the neglect of the acceleration of the field velocity vector
direction in field space.

To construct this model for kination cosmology in the context
of inflation, we will assume the following degrees of freedom:

\begin{enumerate}
\item {\em Inflaton and quintessence}. A real scalar field degree of freedom $\Phi$
plays the role of both the inflaton and quintessence.  Of course, a complete supersymmetric embedding of this scenario would require a complex scalar degree of freedom. We neglect this detail here for
simplicity, as we are first  concerned with generic dynamical settings, and note that there is no straightforward insurmountable obstacle to extending this scenario into a fully supersymmetric framework.

\item {\em MSSM fields}. We assume the presence of electroweak to TeV scale MSSM field degrees
of freedom; the lightest neutralinos are stable LSP's. We will denote the
neutralino LSP as $\chi$ and other generic fields as $\psi$.

\item {\em Couplings}. We assume for simplicity that $\Phi$ is coupled to MSSM fields
only through the minimal gravitational coupling.   Scenarios with additional couplings are of course plausible, but we do not consider them in detail here because the resulting constraints are highly model dependent.   Furthermore, since viable phenomenology requires the quintessence field to be very
weakly coupled to the (MS)SM fields, the minimal gravitational coupling
scenario is {}``natural'' within the quintessence paradigm (of course, quintessence itself has a doubtful status from the point of view of effective field theory). 
\end{enumerate}
For simplicity, we will also functionally tune the potential
({\it i.e.}, assume a specific form of the nonrenormalizable operators). Such
simplifications are reasonable for this first attempt at model
building of this kind, given the current incomplete understanding of
possible UV completions of the MSSM as well as the typical
difficulties of embedding inflation and quintessence in the context of
effective field theories.  Future model-building attempts will need to
address this issue.

Our scenario is predicated on the physical picture that $\Phi$
receives a kick at the end of the inflationary period.  To this end, we consider a scalar field potential with a stepfunction-like behavior, such that $\Phi$ is potential energy
dominated at the top of the step and then becomes kinetic energy
dominated when $\Phi$ drops off the cliff of the step. As an example,
consider the ansatz 
\begin{equation}
V(\Phi)\approx\Omega_{\Lambda}\rho_{c}\left [1+b\cosh^{2}(\lambda\Phi)
\right ]+\left[V_{0} + \beta \ln\left(\frac{(\Phi-\Phi_c)^2}{\mu^2}
+\delta^2 \right) \right]
S(\Phi),\label{eq:potentialwithstep}\end{equation} in which $\beta$ is a
constant energy density scale controlling the slow roll properties of
the period of inflation, $\delta$ is a small constant inserted purely to
regularize the logarithm when $\Phi-\Phi_c$ vanishes, and $S(\Phi)$ is a
steplike function.  An example would be $S(\Phi)=(1- \tanh[\alpha
(\Phi-\Phi_c)])/2$, which is unity for $\alpha(\Phi-\Phi_{c})<1$ and
smoothly goes to zero after $\alpha(\Phi-\Phi_{c})\gg1$ for sufficiently
large values of $\alpha$.  This potential is identical to Eq.
(\ref{eq:potential1}) except with the addition of the $\left[V_{0} +
\beta \ln\left(\frac{(\Phi-\Phi_c)^2}{\mu^2} \right) \right] S(\Phi)$
term responsible for the inflationary period, which in turn shuts off at
$\Phi=\Phi_{c}$.  Note that the constant $\delta$ plays no important
role but to make the logarithmic function regular, and as we will
comment more explicitly later, we can choose a sufficiently small
$\delta$ as not to change any of the inflationary analysis. Hence, it
can be dropped from the anlysis of the inflationary period.

Let us consider the constraints on the scales $V_{0}$, $\alpha$, and
$\beta$ in this scenario (we will set $\mu=\Phi_c$ without any loss of
generality). Inflation occurs for $(\Phi-\Phi_{c})<0$, and ends when
the inflaton receives a hard kick at $\Phi\approx\Phi_{c}$, leading to
kination domination. During the kick at the time $t_e$ of the end of
inflation, the non-adiabatic time variation of the background
gravitational field generates particles with energy density
$\rho_{\psi}(t_{e})$ (a similar reheating scenario was considered in 
\cite{Spokoiny:1993kt}).  The mechanics of this is explained in the
appendix. If the particles that are produced have masses much smaller
than the expansion rate $H_{e}\sim\sqrt{V_{0}/3}/M_{p}$ at the end of
inflation, then \begin{equation} \rho_{\psi}\left (t_{e} \right )\sim 
\frac{\pi^2}{30}g_{*} (T_e \approx H_{e}/2 \pi)\left (\frac{H_{e}}{2\pi}
\right)^{4},
\label{eq:hawkingenergy}\end{equation}
in which 
$g_{*}$ counts the number of light degrees of freedom.  This can be
seen as the situation where the all of the species which couple to the large $\Phi$
vacuum expectation value (VEV) have decoupled and the others are lighter than $H_{e}$.  We have made a special simplifying assumption that there is a large number of light species
despite the large $\Phi$ VEV.  (Indeed, the lack of such light
particles can lead to a moduli problem as the VEV can induce large
masses to the fields to which it couples.  Note that even in regions
where the $\Phi$ VEV is zero, large finite density masses can be
induced for what would otherwise be light fields \cite{Berkooz:2005sf,Berkooz:2005rn}.)
 Due to the kick and the
sudden drop of the potential, the kinetic energy of $\Phi$ just after
inflation ends will be of order $V_{0}$ and will dilute as
$1/a^{6}$. Athough the relativistic species $\psi$ initially is out of
equilibrium, the energy density can be characterized by an approximate
temperature $T_{e}\sim(\rho_{\psi}/g_{*}(T_{e}))^{1/4}$.  As we will
discuss below, $\psi$ will eventually equilibrate and the relativistic
energy density will dilute as $1/a^{4}$.

In this class of scenarios, the initial values of $\rho_\psi$ and $\rho_\Phi$ are interconnected, such that for any given reheating energy density $\rho_\psi$, there is a predicted
value of
$\eta_\Phi \equiv \rho_{\Phi}/\rho_{\gamma} |_{T=1\,{\rm MeV}}$,
which is phenomenologically required to be less than about 1 (at $2
\sigma$) by the time the photon temperature is of order $1$ MeV.
Since $\rho_{\Phi}/\rho_{\gamma}\approx a^{-2}\propto T^{2}$
 during the kination period, $\eta_\Phi$ is given by
\begin{equation} \eta_{\Phi}\sim 33 \frac{(\textrm{1
MeV})^{2}V_{0}}{\rho_{\psi}^{3/2}(T_{e})  \sqrt{g_{*}(T_{e})}}.
\label{eq:etadesired}\end{equation} 
Combining Eqs.~(\ref{eq:hawkingenergy}) and (\ref{eq:etadesired}), we
find that in order to obtain a desired $\eta_{\Phi}$, the inflationary
energy density must be of the form \begin{equation} V_{0}\sim \left (3.9
\times10^{13}\mbox{GeV} \right )^{4}\eta_{\Phi}^{-1/2}\left
(\frac{g_{*}(T_e)}{100} \label{eq:V0sol} \right
)^{-1}.
\end{equation} 
A larger $\eta_{\Phi}$ requires a smaller $V_{0}$, because the radiation energy density at the
end of inflation is proportional to $V_{0}^{2}$, and an increased radiation
energy density corresponds to an increased scale factor growth before the
temperature reaches 1 MeV. 

Eq.~(\ref{eq:V0sol}) is a remarkable result as it sets an approximate
upper bound on $V_{0}$ if a non-negligible $\eta_{\Phi}$ is to be
achieved. Note that this did not depend on the details of the
inflationary model, but only on the fact that a period of kination
domination occurs just following inflation together with reheating.
Although such upper bounds have not been imposed in previous studies
\cite{Salati:2002md,Profumo:2003hq}, this result appears to be quite
generic within a large class of inflationary models.

An interesting ramification of this bound on $V_0$ results from the fact that the detection
of inflationary tensor perturbations in the foreseeable future
requires $V_{0}\gtrsim(3\times10^{15}\mbox{GeV})^{4}$ (corresponding to
a tensor to scalar ratio of about $10^{-4}$) \cite{Knox:2002pe,Kesden:2002ku,Amblard:2006ef,Amarie:2005in}.  Hence, if tensor perturbations are detected, $\eta_\Phi$ must satisfy the following approximate bound:
\begin{equation}
\eta_{\Phi} \lesssim  10^{-13} \left(\frac{\theta}{10}
\right)^2 \left(\frac{r_{\rm min}}{10^{-4}}\right)^{-2}
\left(\frac{g_{*}(T_e)}{100} \right)^{-2}.
\label{eq:exclusion}
\end{equation} 
In the above, $\theta \equiv V(\Phi_N)/V(\Phi_e)$ is the ratio of the potential between the time when the
largest observable scales left the horizon and the time when inflation ends,
and $r_{\rm min}$
is the minimum detectable tensor to scalar ratio defined as $16
\epsilon$ evaluated at the $0.002 \mbox{Mpc}^{-1}$ Hubble crossing scale,
where
$\epsilon \equiv \frac{M_p^2}{2} (V'(\phi)/V(\phi))^2$
is the usual potential expansion slow roll parameter. Therefore, larger values of $\eta_\Phi$ would be ruled out in this scenario if tensor perturbations are measured.  Note that a nonzero
$\eta_\Phi \ll 10^{-6}$ causes an $\mathcal{O}(10^6 \eta_\Phi)$ change
in the prediction of the relic abundance compared to the standard scenario
\cite{Salati:2002md,Profumo:2003hq,cekm07}.  In the event of a positive tensor perturbation measurement, $\eta_\Phi$ would be bounded to be so small that kination dominated scenarios cannot be effective in changing the thermal relic abundance from the values obtained using standard cosmology ({\it i.e.}, the connection between collider measurements and dark energy would be lost).

The constraint for obtaining the required number of efolds is given
by:\begin{equation} N\gtrsim\ln \left (\sqrt{\theta} \frac{H_e}{H_{0}}
\right )+\frac{1}{3}\ln \left
(\frac{g_{*S}(T_{RH})}{g_{*S}(T_{0})}\right
)+\ln\frac{T_{0}}{T_{RH}},\label{eq:efold}\end{equation} in which
$H_{0}$ is the expansion rate today, and in our scenario the  reheating
temperature $T_{RH}\sim H_{e}/2\pi$,  
indicating that the explicit $H_e$ dependence drops out of Eq.~(\ref{eq:efold}).  The
requisite number of efoldings is slightly larger than the usual
inflationary scenario with reheating during coherent oscillations
since there is no corresponding scale stretching that would have
occurred during the coherent oscillations phase.  As there
is no explicit $H_e$ dependence, for $\theta \sim {\mathcal O}(10)$ the required number of efolds of inflation is 
\begin{equation}
N \gtrsim 71,
\label{eq:requiredefolds}
\end{equation} 
nearly independently of the inflationary potential.  Note that in the
absence of an unusual initial state of the vacuum, the number of efolds
required after the onset of inflation to achieve the usual
inflationary predictions is very small \cite{Chung:2003wn}.

Since $V_0$ is fixed by Eq.~(\ref{eq:V0sol}) and $\epsilon$ is
fixed by the density perturbation constraint
\begin{equation}
P_{\mathcal{R}} (0.002 \mbox{ Mpc}^{-1}) = \frac{\theta V_0}{24 \pi^2 \epsilon M_p^4} \approx
2\times 10^{-9},
\end{equation}
$\beta$ is approximately given by
\begin{equation}
\beta \approx (8.3 \times 10^{10} \mbox{GeV})^4 \left( \frac{g_{*S}(T_e)}{100}\right)^{-2} \left(\frac{\theta}{\eta_\Phi} \right).
\end{equation}
For this class of models  $V_0 \gg \beta$, which implies that $\theta \approx 1$ during the
approximate 70 efoldings of interest since the potential depends only logarithmically on $\Phi$ during
inflation.   Nonetheless, 
it is still useful to keep the $\theta$ dependence explicitly to keep track of the model dependence, since
expressions like  Eq.~(\ref{eq:exclusion}) are not strongly dependent
on the shape of the inflationary potential.

To determine the very model dependent field value during which
$k\approx 0.002 \mbox{Mpc}^{-1}$ leaves the horizon (corresponding
to about 68 efoldings before the end of inflation), we use the
usual field integral over $1/\sqrt{2 \epsilon(\Phi)}$, which yields
\begin{equation}
\Phi - \Phi_c \approx- 8 \times 10^{-5} \left(\eta_\Phi \right)^{-1/4}
\left(\frac{g_{*S}(T_e)}{100} \right)^{-1/2}    M_p \ll \Phi_c.
\label{eq:observablescales}
\end{equation}
Eq.~(\ref{eq:observablescales}) implies that unless there is fine tuning of the initial
conditions, the number of efoldings will be much larger than the
requisite number of Eq.~(\ref{eq:requiredefolds}).  Finally, the end
of inflation, which is determined by $|M_p^2 V''(\Phi)/V(\Phi)|\approx 1$ (since afterwards
$\Phi$ rolls quickly to make $\epsilon=1$), is given by
\begin{equation}
\Phi_e-\Phi_c \approx -6\times 10^{-6} \left(\eta_\Phi \right)^{-1/4}
\left( \frac{g_{*S}(T_e)}{100}  \right)^{-1/2} M_p,
\end{equation}
which means that inflation ends very close to $\Phi_c$ (by
construction) and far away from scales of interest of
Eq.~(\ref{eq:observablescales}). This places a constraint
on the steplike function $S$, since the slow roll behavior should not
be disrupted by the slope of this steplike function before the
required number of efoldings of inflation.  For example, if
$S(\Phi)=(1- \tanh[\alpha (\Phi-\Phi_c)])/2$, $\alpha$ is bounded by
\begin{equation}
\alpha \gtrsim \frac{10^6}{M_p} \eta_\Phi^{1/4} \left(
\frac{g_{*S}(T_e)}{100} \right)^{1/6}.
\label{eq:alphacond}
\end{equation}
It is also straightforward to show that in order for $\Phi$ kination domination
to occur at the end of inflation, the potential that kicks $\Phi$ must
satisfy the following condition:
\begin{equation}
-\sqrt{2} M_p \frac{V'(\Phi)}{V(\Phi)} \gtrsim 6,
\end{equation}
a much less stringent constraint than Eq.~(\ref{eq:alphacond}).
Finally, as we have commented near Eq.~(\ref{eq:potentialwithstep}),
we now see explicitly that if we make $\delta \lesssim 10^{-7}$ for
$|\Phi_c|/M_p \gtrsim 10$, $\delta$ does not change the inflationary
analysis.

We have checked the details of this inflationary scenario with
explicit numerical computations.  The analytic discussion above is in
good agreement with the numerical results.
 
\subsection{Inflation to Kination Domination}
The steplike feature in the potential presented in
Eq.~(\ref{eq:potentialwithstep}) suggests an interesting new classical equation of motion which is exactly solvable that is relevant for this class of scenarios.  If gravity is turned off and the $\ln$ term is neglected,
the equation of motion for the coherent state of $\Phi$ at the end of
inflation can be written as\begin{equation} \ddot{\Phi}- \frac{V_0 \alpha}{2}
\mbox{sech}^2 \left(\alpha (\Phi- \Phi_c) \right)=0,\end{equation} in which we
have omitted the negligible contribution proportional to
$\Omega_{\Lambda}\rho_{c}$.  This class of potentials allows us to now
rewrite this equation in the limit $|\alpha \Phi_{c}| \gg 1$ (the
limit of interest according to Eq.~(\ref{eq:alphacond})) as follows:
\begin{equation}
\ddot{\Phi}-V_{0}\delta(\Phi-\Phi_{c})\approx0.\label{eq:approximatedsol}\end{equation}
The parameter $\alpha$ has disappeared from the equation
of motion; this is not surprising given the stepfunction-like behavior
of the potential in this limit. The beauty of
Eq.~(\ref{eq:approximatedsol}) is that it can be solved exactly.  The
solution is given by \begin{equation}
\Phi=\Theta(t_{c}-t)\{\Phi_{i}+(t-t_{i})\dot{\Phi}_{i}\}+\Theta(t-t_{c})\{\Phi_{c}+(t-t_{c})\gamma\},\end{equation}
where $\Theta(x)$ is a stepfunction,
$\gamma\equiv\sqrt{2V_{0}+\dot{\Phi}_{i}^{2}}$, $\Phi_{i}$ and
$\dot{\Phi}_{i}$ are initial values of the field and its velocity at
time $t_{i}$, and $t_{c}$ is the time at which $\Phi$ reaches
$\Phi_{c}$. Therefore, since the initial time variation of
the field $\dot{\Phi}_{i}$ is small compared to $\sqrt{V_{0}}$ , the
field obtains a strong kick at $t=t_{c}$ to obtain the final state
velocity of $\dot{\Phi}\sim\sqrt{2V_{0}}$.

\subsection{Does Dark Matter Ever Reach Equilibrium?}

Another difference from more traditional inflationary scenarios is that the dark matter is initially out of thermal equilibrium after reheating (or more accurately, entropy production) in
this scenario. To see this, first consider the more traditional inflationary paradigm. The neutralino $\chi$ self-annihilation rate behaves as\begin{equation}
\Gamma_{\chi}\sim\langle\sigma v\rangle n_{\chi}^{eq}\sim\alpha^{2}T,\end{equation}
where $\chi$ is relativistic, and  $\alpha\equiv g_{W}^{2}/(4\pi)$
is the weak coupling expansion parameter. On the other hand,  the expansion rate after reheating in standard inflationary scenarios behaves as\begin{equation}
H\sim\sqrt{g_{*}}\frac{T^{2}}{M_{p}},\end{equation}
which means that the neutralinos are in equilibrium as long as \begin{equation}
T\lesssim2.6\times10^{14}\mbox{GeV}\left (\frac{\alpha}{1/30} \right )^{2}\left (\frac{g_{*}}{100} \right)^{-1/2}.\end{equation}
In our scenario, however, $H$ is governed by $\dot{\Phi}$ (not $T$) during kination
domination, which is also when $\chi$ freezes out:\begin{equation}
H\approx\sqrt{\frac{V_{0}}{3M_{p}^{2}}}\left(\frac{a_{e}}{a}\right)^{3}\approx\sqrt{\frac{V_{0}}{3M_{p}^{2}}}\left(\frac{2\pi T}{H_{e}}\right)^{3}.\end{equation}
In the above, we have used the fact that $\dot{\Phi}^{2}\propto1/a^{6}$ during
kination domination. Hence, the neutralinos reach equilibrium through
self-annihilations only for 
\begin{equation}
T \lesssim 8 \times  10^{5} \mbox{GeV}
\left(\frac{\alpha}{1/30}\right) \left (\frac{V_{0}}{[3.9
    \times10^{13}\mbox{GeV}]^{4} }\right )^{1/2}.
\label{eq:temperaturebound}
\end{equation}
 Comparing this temperature with the original relativistic species
temperature of\begin{equation}
\frac{H_{e}}{2\pi}\approx\left(\frac{V_{0}}{[3.9 \times10^{13}\mbox{GeV}]^{4}}\right)^{1/2} 6 \times10^{7}\mbox{ GeV},\end{equation}
we see that there is a long period in which the neutralino self-annihilations
are out of equilibrium after the entropy in the universe has been
produced. The fact that Eq.~(\ref{eq:temperaturebound}) exists for
$T\gg 1$ GeV is important, since our goal is to embed the modified dark matter freeze-out
scenario in which the dark matter was originally in equilibrium.

%

\section{\label{sec:otherpredictions} Other Predictions for Cosmology}

One of the main advantages of embedding the kination domination
scenario within an inflationary cosmological setting is that other
predictions for observables can be made whose experimental
confirmation would either support or rule out the scenario.  We have already
seen that if tensor perturbations are measured in the near future,
this scenario is ruled out.  There are many other observables
correlated with this dark matter scenario.  Although an exhaustive
analysis of these signatures is beyond the scope of this paper, we
briefly discuss several possibilities here.

In the present scenario, an out of equilibrium effective ``temperature'' scale as
high as $10^8 \eta_\Phi^{-1/4}$ GeV is reached at the end of inflation
and the equilibration of the (MS)SM particles occurs by temperatures of about $10^6 \eta_\Phi^{-1/4}$ GeV.  During the
time until the temperature is below $\sim 1$ GeV, the Hubble expansion rate
differs significantly from the usual radiation dominated universe
value.  This implies that any physics which depends both on the temperature and
the Hubble expansion will be modified.

One testable example is the gravity wave production during the
electroweak phase transition (see for example
\cite{Kosowsky:1992rz,Kosowsky:1991ua,Apreda:2001us,Dolgov:2002ra,Nicolis:2003tg,Grojean:2006bp}),
where the peak frequency of the gravity waves are set by the Hubble
scale.  Since $H$ is about $10^5 \sqrt{\eta_\Phi}$ of the usual Hubble
value (for the same temperature), the effects can be large even for a
very small $\eta_\Phi$.  Exact details will require a careful
reanalysis of the gravity wave production.  The effects on electroweak
baryogenesis are expected to be weaker since the out of equilibrium
condition is primarily provided by the bubble wall velocity \cite{Joyce:1997fc,Servant:2001jh}.  A more
careful investigation of this issue is left for future work.

Another prediction of this scenario is that since the temperature at
which equilibrium is reached for any heavy lepton number carrying particle
(such as a right handed neutrino) will be relatively low as in
Eq.~(\ref{eq:temperaturebound}), possible leptogenesis mechanisms in
this context will necessarily be either nonthermal or nonstandard.  It will be
interesting to explore what kinds of leptogenesis scenarios are
viable for this class of models. 

There are also predictions associated with particle astrophysics.  For example, it
is well known that explaining the HEAT measurement (and other cosmic
ray positron measurements) of excess positrons around and above 7 GeV
requires an efficient annihilation of neutralinos (or other thermal relics in the context of models of extra dimensions, etc.) within our halo \cite{Kane:2002nm,Kane:2001fz,Hooper:2004xn}.  One of the many problems associated
with this efficient annihilation scenario is that the relic abundance
of neutralinos is generically too low (by a factor of 10 to 100) to explain
most of the dark matter energy within the context of standard thermal
scenarios.  The kination scenario can clearly give the
necessary boost to resurrect the neutralino dark matter annihilation
explanation of the excess positrons.  There are also other cosmic ray
signatures which may shed light on nonstandard $H$ behavior \cite{Donato:2006af,Schelke:2006eg}.

Finally, there are other possible signatures such as the change
in BBN due to the effects of residual annihilations after freeze-out
\cite{Jedamzik:2004ip}, and the change in cosmic string generated
gravity wave signature \cite{Damour:2004kw,Ringeval:2005kr} due to the
change in $T/H$ involved in the scaling behavior of cosmic strings.
Since much of cosmology is about studying the out of equilibrium
phenomena generated by the expansion of the universe within a finite
temperature setting, further signatures related to a non-standard
relationship between $T$ and $H$ will appear as we learn more about the early history of our universe.

\section{Conclusions}

The possibility of a new scalar is generic in extensions of physics
beyond the standard model of particle physics and cosmology
(particularly in those containing a dilatonic field degree of
freedom).  It is natural to expect that the dark energy density is
connected with such a new scalar field degree of freedom. However, the
elusive nature of dark energy, and its requisite small couplings to
observable fields, make such conjectures difficult to verify or
disprove.

We have constructed a viable class of inflationary scenarios which
exhibit a period of kination domination after inflation where the
inflaton plays the role of the quintessence.  Such scenarios
have the intriguing feature that they can lead to observable consequences for the dark matter freeze-out of thermal relics expected in many TeV-scale extensions of the SM, which can be
tested at the LHC and ILC.  We have focused here on supersymmetric scenarios as prototype examples, for which the connections between astroparticle and collider physics have been extensively explored \cite{Polesello:2004qy,Battaglia:2004mp,%
Allanach:2004xn,Battaglia:2004gk,Bourjaily:2005ax,Belanger:2005jk,Moroi:2005nc,%
Birkedal:2005jq,Chattopadhyay:2005mv,Battaglia:2005ie,Moroi:2005zx,Birkedal:2005aa,%
Nojiri:2005ph,Baltz:2006fm,Djouadi:2006pg,White:2006wh,Carena:2006gb,Cirigliano:2006dg}. 
However, other WIMP candidates have recently emerged
in models with flat \cite{Servant:2002aq,Cheng:2002ej} or warped \cite{Agashe:2004ci,Agashe:2004bm}
extra dimensions, in Little Higgs theories 
\cite{Birkedal-Hansen:2003mp,Cheng:2003ju,Katz:2003sn,Cheng:2004yc,Birkedal:2006fz}, or in technicolor models \cite{Kainulainen:2006wq}, to which the cosmological scenarios described in this paper could also be applied.

The advantage of embedding kination-dominated quintessence models
within an inflationary context is that it allows for many other
correlated cosmological predictions which can corroborate or rule out
such kination dominated scenarios.  The most robust, nearly model
independent, signature of this class of models is the absence of
measurable tensor perturbations, such that any positive detection of
tensor perturbations in upcoming experiments can rule out this
scenario (at least as far generating an observable shift in the DM
abundance is concerned).  Other examples are measurements of gravity
waves from the electroweak phase transition, shifts in the predictions
for baryogenesis/leptogenesis, implications for the cosmic ray flux
from dark matter annihilations, and other phenomena which depend on
the ratio of the photon temperature to the Hubble expansion rate
$T/H$.

The class of models in this paper is meant to be illustrative and
represent early attempts at model-building, and as
such certain features are not optimal. For example, one might argue
that the toy model presented here may not be easily achievable from an
effective field theory point of view. However, it is reasonable to
believe that this class of models are as potentially viable as any quintessence and
most inflationary models that are considered seriously  
in the current literature.  Given that such scenarios may be testable through
their potentially dramatic interconnection with dark matter
predictions and TeV-scale particle physics, they represent an
intriguing and potentially fruitful ground for quintessence
model-building which warrants further exploration.
\begin{acknowledgments}
We thank E.~Chun, G.~Kane, and G.~Servant for useful conversations.
The work of DJHC was supported by DOE Outstanding Junior Investigator
Program through Grant No.  DE-FG02-95ER40896.  The work of KTM was supported by DOE Outstanding Junior Investigator Program through Grant No.  DE-FG02-97ER41029. 
\end{acknowledgments}

\begin{appendix}
\section{Particle Production Computation}
In this Appendix, we consider the particle production computation for a real scalar degree
of freedom coupled to gravity.  We start with the perturbative expression
for the Bogoliubov coefficient as an integral over conformal time \cite{Chung:1998bt}:\begin{eqnarray}
\beta_{k} & \approx & \int d\eta\frac{w'_{k}}{2w_{k}}\exp[-2i\int w_{k}d\eta]\nonumber \\
 & = & \int\frac{d(w_{k}^{2})}{4w_{k}^{2}}\exp[-i\int\frac{d(w_{k}^{2})}{w_{k}'}],\label{eq:phasefactor}\end{eqnarray}
where \begin{equation}
w_{k}^{2}\approx k^{2}+(\frac{1}{6}+\xi)Ra^{2},\end{equation}
which assumes that the effective mass is dominated by the Ricci scalar
$R\gg m_{\chi}^{2}$ (note that $\xi=0$ corresponds to minimal gravitational
coupling). Using the approximation\begin{equation}
H^{2}=\left\{ \begin{array}{cc}
V_{0}/(3M_{p}^{2}), & \eta<\eta_{c}\\
V_{0}(a_{e}/a)^{6}/(3M_{p}^{2}), & \eta\geq\eta_{c}\end{array}\right .\end{equation}
 we find\begin{equation}
w_{k}^{2}\approx-6(\frac{1}{6}+\xi)a^{2}\left(\frac{2V_{0}}{3M_{p}^{2}}\Theta(\eta_{c}-\eta)-\frac{V_{0}}{3M_{p}^{2}}\left(\frac{a_{e}}{a}\right)^{6}\Theta(\eta-\eta_{c})\right)+k^{2},\end{equation}
\begin{equation}
w_{k}'\approx\frac{-1}{2w_{k}}\frac{|w_{k}^{2}-k^{2}|^{3/2}}{\sqrt{\frac{1}{6}+\xi}}\left[\frac{1}{\sqrt{3}}\Theta(\eta_{c}-\eta) +2\sqrt{\frac{2}{3}}\Theta(\eta-\eta_{c})\right],\end{equation}
in which $\Theta(z)$ is a unit stepfunction which evaluates to $1$
for $z>0$. 
For the limits of the integral over $w_{k}^{2}$ in Eq. (\ref{eq:phasefactor}), we have\begin{equation}
-(\frac{1}{6}+\xi)a_{e}^{2}\frac{4V_{0}}{M_{p}^{2}}+k^{2}\leq w_{k}^{2}\leq k^{2}\end{equation}
during inflation, and\begin{equation}
k^{2}\leq w_{k}^{2}\leq(\frac{1}{6}+\xi)a_{e}^{2}\frac{2V_{0}}{M_{p}^{2}}+k^{2}\end{equation}
 after the end of inflation. The conformal time $\eta$ is not a single
valued function of $w_{k}^{2}$ at the transition at the end of inflation.
However, since the non-single valued time period is arbitrarily short
(in the limit that the potential behaves like a step function), this time period can be excised
from the computation without loss of numerical
accuracy, as long as a UV cutoff is imposed. The reason for the UV cutoff 
is that the peak strength of the nonadiabaticity responsible for particle
production is precisely determined by the detailed gravitational dynamics
of the transition time period, which we excise to simplify the
computation. Since the particle production is through gravitational
curvature, there will generically be an exponential
cutoff in momentum of the particles produced at $H_{e}$, the expansion
rate at the end of inflation. 
With this simplification, 
Eq. (\ref{eq:phasefactor}) can be written as\begin{eqnarray}
\beta_{k} & \approx & \frac{1}{4}\int_{k^{2}}^{x_{1}}\frac{dx}{x}\exp\left[-i 2\sqrt{3} \sqrt{\frac{1}{6}+\xi} \left(\frac{2\sqrt{x}\sqrt{k^{2}-x}}{x-k^{2}}-2\arctan\left[\frac{\sqrt{k^{2}-x^{2}}\sqrt{x}}{x-k^{2}}\right]+C_{1}(k^{2})\right)\right]\nonumber \\
 &  & +\frac{1}{4}\int_{x_{2}}^{k^{2}}\frac{dx}{x}\exp\left[-i\sqrt{\frac{3}{2}}\sqrt{\frac{1}{6}+\xi} \left(\frac{-2\sqrt{x}}{\sqrt{x-k^{2}}}+2\ln\left[\sqrt{x-k^{2}}+\sqrt{x}\right]+C_{2}(k^{2})\right)\right],\label{eq:betaclose}\end{eqnarray}
in which $C_{1,2}$ are constant phase factors which  are independent of the integration variable $x$ and depend on $k^{2}$,
while \begin{eqnarray}
x_{1}&\equiv& k^{2}-\frac{4V_{0}}{M_{p}^{2}}(\frac{1}{6}+\xi)a_{e}^{2}\\
x_{2}&\equiv& k^{2}+\frac{2V_{0}}{M_{p}^{2}}(\frac{1}{6}+\xi)a_{e}^{2},\end{eqnarray}
which correspond to $w_{k}^{2}$ just before and after the end of
inflation (approximately a step function transition).

To obtain an estimate for $|\beta_{k}|^{2}$, we will neglect the
interference term in $|\beta_{k}|^{2}$, which means that we can
neglect the integration constants $C_{i}$.  We will also utilize the
fact that the contribution to the $dx$ integral in
Eq.(\ref{eq:betaclose}) is appreciable only when $x$ is near $x_i$ and
far away from $k^2$ because of the damping phase oscillations near
$x\approx k^2$. Finally, we will only account for contributions with
$w_{k}^{2}>0$, so as to maintain the particle interpretation of the
massless modes produced.  We then find
\begin{equation}
|\beta_{k}|^{2}\sim\frac{1}{16}\left[\left(\ln \left [1-\frac{4V_{0}}{M_{p}^{2}}\left (\frac{1}{6}+\xi \right)\frac{a_{e}^{2}}{k^{2}}\right ]\right)^{2}+\left(\ln \left [1+\frac{2V_{0}}{M_{p}^{2}}\left (\frac{1}{6}+\xi \right )\frac{a_{e}^{2}}{k^{2}}\right ]\right)^{2}\right]\Theta(k-k_{min})\Theta(aH_{e}\tilde{\lambda}-k),\end{equation}
 where $k_{min}=\sqrt{4V_{0}/(M_{p}^{2})(1/6+\xi)}$
is an appropriate infrared cutoff imposed to maintain the particle
production interpretation of the massless modes, and $aH_{e}\tilde{\lambda}$
is the UV cutoff,  with $\tilde{\lambda}\sim\mathcal{O}(1)$
reflecting the uncertainty in the UV cutoff function.

Taking the approximation that the log factors contribute
$\mathcal{O}(1)$, the energy density of the real scalar degree of
freedom that is produced is\begin{equation} \rho \sim
\frac{1}{a^{3}}\int\frac{d^{3}k}{(2\pi)^{3}}\left (\frac{k}{a}\right
)|\beta_{k}|^{2} \sim \frac{\tilde{\lambda}^{4}}{32\pi^{2}}H_{e}^{4} =
\frac{\tilde{\lambda}^{4}\pi^{2}}{2}\left (\frac{H_{e}}{2\pi}\right
)^{4}.\end{equation} Taking the parameterization of the thermal
equilibrium abundance of one real scalar degree of freedom as
$\rho=(\pi^{2}/30)T_{{\rm eff}}^{4}$, where $T_{\rm eff}$ is the
effective temperature and $g_{*}$ counts the number of degrees of
freedom, we find\begin{equation} T_{{\rm
eff}}\sim\frac{H_{e}}{2\pi}(2\tilde{\lambda}).\end{equation} For an
order of magnitude estimate, we will absorb the uncertainty factor
$2\tilde{\lambda}$ in the uncertainty in the effective number of
degrees of freedom during the {}``reheating'' stage at the end of
inflation, and therefore write\begin{equation}
T_{e}\sim\frac{H_{e}}{2\pi}\end{equation} throughout the paper.
\end{appendix}
\end{document}